\documentclass[12pt,a4paper]{article}
\title{Reconstructing the density matrix of a spin $s$ through 
       Stern-Gerlach measurements (II)}
\author{Jean-Pierre Amiet and Stefan Weigert \\
Institut de Physique, Universit\'e de Neuch\^atel\\
Rue A.-L. Breguet 1, CH-2000 Neuch\^atel, Switzerland\\
\tt stefan.weigert@iph.unine.ch}
\date{March 1999}
\newcommand\be{\begin{equation}}
\newcommand\ee{\end{equation}}
\newcommand\ket[1]{|#1\rangle}
\newcommand\bra[1]{\langle #1|}
\newcommand\braket[2]{\langle #1|#2\rangle}
\begin{document}
\maketitle
\begin{abstract}
The density matrix of a spin $s$ is fixed uniquely if the probabilites to obtain the value $s$ upon measuring $\vec{n} \cdot \vec{\bf s}$ are known for $4s(s+1)$ appropriately chosen directions $\vec{n}$ in space. This non-redundant set of numbers are just expectation values of the density operator in coherent spin states, and they are easily obtained in an experiment carried out with a Stern-Gerlach apparatus. Furthermore, the experimental data can be inverted {\em explicitly} which is necessary for a parametrization of the statistical operator by the $4s(s+1)$ positive parameters. 
%
\end{abstract}
If an infinite ensemble of identically prepared quantum systems is available, 
the rules of quantum mechanics allow one to extract complete information about the quantum state. The required inversion, expressing the statistical operator entirely in terms of measurable quantities, is not straightforward from a mathematical point of view, and the data acquisition in the laboratory is challenging since it is neccessary to deal reliably with individual quantum systems. 

The experimental verification of a state reconstruction scheme requires high standards in the preparation of individual quantum systems. By now, these standards have been met for quantum systems such as an electromagnetic wave \cite{smithey+93}, vibrating molecules \cite{dunn+95}, ions caught in a trap \cite{leibfried+96}, and atoms moving freely in space after scattering from a double slit \cite{kurtsiefer+97}. State reconstruction is reviewed in  \cite{raymer97} and \cite{leonhardt97}, for example, both from the theoretical and experimental 
point of view.

State reconstruction for a quantum system with a finite number of states appears to be slightly easier than for particle systems living in a Hilbert space with countably infinite dimension.
Various answers to the problem have been obtained for both mixed and pure states of 
spins with length $s$. {\em Pure} states in two- or three-dimensional Hilbert spaces 
($s=1/2,1$) have been treated in a straightforward way while the general case is more complicated \cite{weigert92,amiet+99/1}. Using Feynman filters, a phase sensitive version of a Stern-Gerlach apparatus \cite{feynman+65}, one can determine directly moduli and (relative) phases of the individual matrix elements of the density operator \cite{gale+68} describing a {\em mixed} spin state. As shown in \cite{band+71a}, the expectations of $4s(s+1)$ linearly independent spin multipoles 
fix a unique density operator; however, no method has been indicated how to 
determine experimentally these values.
An experimentally more realistic approach has been proposed in \cite{newton+68}: the density matrix is determined if $2s(4s+1)$ real numbers are measured by using a Stern-Gerlach apparatus oriented along various directions in space. In \cite{amiet+98/1}, a reduction to the minimum number of $4s(s+1)$ measured probabilities has been proposed which, as shown in the Appendix, unfortunately is erroneous.   Alternatively, 
a tomographic approach adapted to finite-dimensional Hilbert spaces allows for experimental reconstruction of quantum states \cite{leonhardt95}. As is common
for methods involving Wigner functions, the information to be extracted from experiments 
is {\em redundant}. For an application of this approach, see \cite{walser+96}, where the  determination of a single quantized cavity mode is treated.      

In this letter, a simple and constructive reconstruction scheme for the mixed state of a spin $s$ is presented in accordance with the following two natural constraints taken from \cite{amiet+98/1}:
\begin{itemize}
\item[(1)] 
     the measurements are performed with a standard Stern-Gerlach apparatus only;
\item[(2)]
     no redundant information is acquired.
\end{itemize}
These conditions guarantee that a standard experimental setting is sufficient to perform 
the reconstruction of mixed spin states in the most economical way.
 
 
The states of a spin of magnitude $s$ belong to a Hilbert space  $\mathcal{H}_s$ of complex dimension $(2s+1)$, carrying an irreducible representation of the group $SU(2)$. The components of the spin operator $\vec {\bf S} \equiv \hbar \vec {\bf s}$ with standard commutation relations $[{\bf s}_x,{\bf s}_y]= i{\bf s}_z, \ldots$, generate rotations about the corresponding axes. The standard basis of the space $\mathcal{H}_s$ is given by the eigenvectors of the $z$ component $ {\bf s}_z = \vec{n}_z\cdot\vec{\bf s} $ of the spin, which are denoted by $\ket{\mu,
\vec{n}_z}, -s\leq\mu\leq s$. The phases of the states are fixed by the transformation under 
the anti-unitary time reversal operator $T$:  $T\ket{\mu,\vec{n}_z}=(-1)^{s-\mu}\ket{-\mu,\vec{n}_z}$, and the ladder operators ${\bf s}_{\pm}= {\bf s}_x {\pm} i {\bf s}_y$ act as usual in this basis:
\begin{equation}
{{\bf s}}_ {\pm} \ket{\mu,\vec{n}_z}
            = \sqrt{s(s+1)-\mu(\mu\pm 1)} \ket{\mu \pm 1,\vec{n}_z} \, .
\label{splusaction}
\end{equation}
The complexified algebra $\mathcal{A}_s$ of {\em observables} in the space $\mathcal{H}_s$ has complex dimension $(2s+1)^2$. It consists of all polynomials in the operators ${{\bf s}}_x$, ${{\bf s}}_y$ and ${{\bf s}}_z$ with complex coefficients and of degree $2s$ at most. A monomial of a degree higher than $2s$ can always be expressed as a linear combination of monomials of lower degree. 

Consider the eigenstates of the operator $\vec{n} \cdot \vec{\bf s}$, 
\be
\vec{n} \cdot \vec{\bf s} \, \ket{ \mu, \vec{n}} = \mu \, \ket{ \mu,\vec{n}} \, , 
\qquad -s \leq \mu \leq s \, ,
\label{Sneigest}
\ee
where the unit vector $\vec{n}= ( \sin \theta \cos \varphi,$ $\sin \theta \sin \varphi, \cos \theta)$, 
$0 \leq \theta \leq \pi, 0 \leq \varphi < 2\pi$, defines a direction in space.
Given a state with density matrix $\rho$, the probability $p_\mu (\vec{n})$ to measure the value $\mu$ with a Stern-Gerlach apparatus oriented along $\vec{n}$ is given by 
\be
p_\mu (\vec{n}) = \mbox{ Tr } [ \, \rho P_\mu ( \vec{n} ) \, ] 
                = \bra{ \mu , \vec{n} } \, \rho \, \ket{ \mu, \vec{n}} \, ,
\label{prob}
\ee
with the projector $P_\mu ( \vec{n} ) = \ket{ \mu, \vec{n}}\bra{ \mu , \vec{n} }$.
For the state with maximal weight, $\mu = s$, the probability $p_s (\vec{n})$ coincides
with the expectation value of $\rho$ in a (standard)\footnote{Other families of coherent states are obtained if a `fiducial' state different from $\ket{s,\vec{n}_z}$ is used  \cite{perelomov86}.} {\em coherent} state \cite{arecchi+72},
\be
\ket{s, \vec{n} } 
    = \exp [ -i \, \theta \, \vec{m}(\varphi) \cdot \vec{\bf s} \, ] \, \ket{s,\vec{n}_z} \, ,
\label{defcs}
\ee
where $\vec{m}(\varphi) = (- \sin \varphi,\cos\varphi,0)$. In other words, the state $\ket{ \vec{n} }$ is the result of rotating the state $\ket{s,\vec{n}_z}$ 
about the axis $\vec{m}(\varphi)$ in the $xy$ plane by an angle $\theta$. It is convenient to combine $(\theta,\varphi)$ into a single complex variable, $z= \tan(\theta/2)\exp [ i \varphi ]$. This corresponds to a stereographic projection of the surface of the sphere to the complex plane. In terms of $z$, a coherent state has the expansion \cite{amiet+91}
\be
\ket{s, \vec{n} }  
     = \frac{1}{(1+|z|^2)^s}\sum_{\mu= -s}^{s}
           \left(
                  \begin{array}{c}
                      2s \\ 
                     s-\mu 
                  \end{array}
           \right)^{1/2} 
z^{s-\mu}\ket{\mu, \vec{n}_z} \, .
\label{expandcs}
\ee
According to (\ref{prob}) the mean value of $\rho$ in this state is 
\be
p_s( \vec{n} ) = \sum_{\mu , \mu' = -s}^s 
                   \left(
                       \begin{array}{c}
                           2s \\ 
                          s-\mu' 
                       \end{array}
                   \right)^{1/2}
                   \left(
                        \begin{array}{c}
                            2s \\ 
                           s-\mu 
                        \end{array}
                   \right)^{1/2}
\frac{\overline{z}^{s-\mu'}z^{s-\mu}}{(1+|z|^2)^{2s}} 
 \rho_{\mu' \, \mu} \, ,
\label{meanvalue}
\ee
where $\overline{z}$ denotes the complex conjugate of $z$. Note that Eq.\ (\ref{meanvalue}) defines a {\em linear} relation between the quantities $p_s( \vec{n} )$ and the unknowns $\rho_{\mu'\, \mu}$. Due to the hermiticity of the density matrix, $\rho_{\mu'\, \mu} = \overline{\rho}_{\mu\, \mu'}$, it contains $(2s+1)^2$ free real parameters (the normalization condition $ \mbox{ Tr } [ \, \rho \, ] = 1 $ will be suppressed for the moment).  
Therefore, the probabilities $p_s( \vec{n} )$ must be known for at least $(2s+1)^2$ 
points $z_{\lambda}$, $\lambda = 1,2, \ldots , (2s+1)^2$. Each of the points $z_\lambda$ in the complex plane corresponds to a point on the sphere or, equivalently, to one spatial direction 
$\vec{n}_\lambda \equiv \vec{n} (\theta_\lambda,\varphi_\lambda)$.  

If the points $z_{\lambda}$ are chosen appropriately, the $(2s+1)^2$ measurable numbers $p_s( \vec{n}_\lambda)$ contain all the information needed to determine the matrix  elements $\rho_{\mu'\, \mu}$ and thus the quantum state. To show this, let us rescale both the measured probabilities,
\be
\widetilde{p}_{\lambda} =  (1+|z_{\lambda}|^2)^{2s} \, p_s( \vec{n}_\lambda) \, ,
\label{scale}
\ee
and the matrix elements of the statistical operator,
\be
\widetilde{\rho}_{k'k} 
= \left(\begin{array}{c}
            2s \\ 
             k' 
         \end{array}
  \right)^{1/2}
  \left(
        \begin{array}{c}
           2s \\ 
            k 
         \end{array}
  \right)^{1/2}
     \rho_{s-k' \, s-k} \, ,
\label{newrho}
\ee
which, for convenience, have been relabelled with nonnegative integers $k=s-\mu$ and $k'=s-\mu'$, $k,k' = 0,1, \ldots, 2s$. Writing down Eq.\ (\ref{meanvalue}) for $(2s+1)^2$ different (as yet unspecified) directions $\vec{n}_\lambda$, one obtains
\begin{equation}
\widetilde{p}_{\lambda} 
         = \sum_{k,k' = 0}^{2s}\overline{z}_{\lambda}^{k'} z_{\lambda}^{k} \, 
                                           \widetilde{\rho}_{k' \, k} \, ,
         \qquad \lambda =1,2,\ldots,(2s+1)^2 \, .
\label{basis-equ}
\end{equation}
It it not obvious how to invert directly the $(2s+1)^2\times(2s+1)^2$ matrix ${\sf N}_{\lambda (k'k)}$ $ \equiv \overline{z}_{\lambda}^{k'} z_{\lambda}^{k}$, which would provide an immediate solution of the problem. By a Fourier transform, however, one can divide the $(2s+1)^2$ coupled equations into smaller sets of equations each of which contains $(2s+1)$ unknowns. For integer spin, $(s+1)$ such sets will emerge while for half integer spin their number is $(s+3/2)$. Select $(2s+1)^2$ directions $\vec{n}_\lambda \equiv$ $\vec{n}_{qr}$ with  
\be
z_{\lambda} \equiv z_{qr} = R_q \exp [i \varphi_{qr}  ] \, , 
                      \qquad 0\leq q,r\leq 2s \, , 
\label{polarz}
\ee
with positive numbers $R_q >0 $, $R_q \neq R_{q'}$ if $q\neq q'$, and 
\be
\varphi_{qr}= \frac{2\pi}{2s+1} ( r + q \Delta  ) \, ,
              \quad 0 \leq \Delta \leq \frac{1}{2s+1} \, .   
\label{points}
\ee
The numbers $R_q$ define $(2s+1)$ circles in the complex plane which correspond to $(2s+1)$ circles on the sphere about the $z$ axis. The values of the angles  $\varphi_{qr}$ define $(2s+1)$ directions equidistant on each circle. An angle $\Delta\neq 0$ provides a shift of the directions on one circle relative to those on the others. It turns out that a nonzero $\Delta$ is only necessary if one deals with the {\em fermionic} problem of state reconstruction (half integer spin) while it can be dropped in the {\em bosonic} case (integer spin).  

Using 
\be
\frac{1}{2s+1}\sum_{r= 1}^{2s+1} \exp [ i(m+k-k')\varphi_{qr}] 
 = \delta_{k'\,k+m} \, + \, \exp [ i 2\pi q \Delta ] \delta_{k' \, k+m-(2s+1)} \, ,  \quad 
   0 \leq m \leq 2s \, , 
\label{krone}
\ee
Eq.\ (\ref{basis-equ}) turns into
\begin{equation}
\widetilde{p}_{qm} 
  = \sum_{k= 0}^{2s-m}R_q^{2k+m} \, \widetilde{\rho}_{k+m \, k} 
         + \exp [i 2 \pi q \Delta ] \, 
           \sum_{k= 2s-m + 1}^{2s}  R_q^{2k+m-(2s+1)} \, \widetilde{\rho}_{k+m-(2s+1) \, k}  \, ,
\label{clean-equ}
\end{equation}
where the shorthand 
\be
\widetilde{p}_{qm} \equiv \frac{1}{2s+1}\sum_{r= 1}^{2s+1}
               \exp [ i m \varphi_{qr} ] \, \widetilde{p}_{qr}
\label{transform}
\ee
has been introduced. A matrix notation will be useful here. Collect the unknowns
associated with a fixed value of $m$ into a vector with $(2s+1)$ components,
\be
\vec{\rho}_m = (\widetilde{\rho}_{m \, 0},\widetilde{\rho}_{1+m \, 1},\ldots,
             \widetilde{\rho}_{2s \, 2s-m};  \widetilde{\rho}_{0 \, 2s-m+1} , \ldots ,
             \widetilde{\rho}_{m-1 \, 2s})
\, , \quad 0\leq m \leq 2s\, ,
\label{unknowns}
\ee
(with no entry on the right of the semicolon if $m=0$) and similarly the data $p_{qm}$ into vectors 
\be
\vec{p}_m = (\widetilde{p}_{1 \, m},\widetilde{p}_{2 \, m}, \ldots ,  
                         \widetilde{p}_{2s+1 \, m} ) \, , \quad 0\leq m \leq 2s\, .
\label{data}
\ee
Then, the relations (\ref{clean-equ}) take the form
\be
\vec{p}_m = {\sf M}_m \, \vec{\rho}_m \, , \quad 0\leq m \leq 2s\, ,
\label{matrixform}
\ee
where the elements $({\sf M}_m)_{qk}$ of the $(2s+1)$ matrices ${\sf M}_m$ with dimension 
$(2s+1) \times (2s+1)$ can be read off from (\ref{clean-equ}). For each value of $m$, Eq.\ (\ref{matrixform}) is  a closed set of equations for the unknowns $\vec{\rho}_m$. If $m$ equals zero, there are $(2s+1)$ real equations for the $(2s+1)$ {\em real} diagonal elements $\vec{\rho}_0$ of the statistical operator. If $m$ equals 1, one has $(2s+1)$ linear complex equations for $(2s+1)$ {\em complex} numbers $\vec{v}_1$. The remaining sets of equations for $m= 2,3,\ldots , 2s$ have the same structure. However, not all of them are independent. In the {\em bosonic} case, the sets come in $s$ pairs: taking the complex conjugate of Eq.(\ref{matrixform}) with label $m_0$, one obtains the equation with index $(2s-m_0)$. Hence, the total number of real independent equations is at most $1 \cdot (2s+1) + s \cdot 2(2s+1) = (2s+1)^2$ which coincides exactly with the number of real free parameters of the density matrix $\rho$. In the {\em fermionic} case, there is again one set ($m=0$) of real equations for the $(2s+1)$ diagonal elements of the statistical operator, plus ($s-1/2$) paired sets of complex equations, plus one `self-conjugate'
set of equations with only $(2s+1)$ real unknowns. This gives a total of $ 1 \cdot (2s+1) + (s-1/2) \cdot 2 (2s+1) + 1\cdot (2s+1) = (2s+1)^2$, as before. Thus, it remains to show that the matrices 
${\sf M}_m$ are invertible for the relevant values of $m$.      

The choice 
\be 
R_q = r^{s-q} \, , \qquad r > 0 \, , 
\label{choice}
\ee
turns each ${\sf M}_m$ into a matrix of type Vandermonde. The {\em bosonic} case $(\Delta = 0)$ is particularly simple: defining 
\be
r_k = \left\{ \begin{array} {ll}
              r^{2k+m} & \mbox{ if }      0 \leq k\leq 2s-m \, , \nonumber \\
              r^{2k+m-(2s+1)} & \mbox{ if } 2s-m+1 \leq k \leq 2s \, , \nonumber
              \end{array}
      \right.
\label{powers}
\ee  
one obtains: 
\begin{equation}
{\sf M}_m = \left(
\begin{tabular}{ccc}
$(r_0)^s$ & $\cdots$ & $(r_{2s})^s$ \\
$\vdots$ & {} & $\vdots$\\
1 & $\cdots$ & 1 \\
$\vdots$ & {} & $\vdots$ \\
$(r_0)^{-s}$ & $\cdots$ & $(r_{2s})^{-s}$
\end{tabular}
    \right) \, . 
\label{fund-matrix}
\end{equation}
The determinant of a Vandermonde matrix is known explicitly \cite{greub63}, leading here to:
\be
\det {\sf M}_m = \left( \prod_{k=0}^{2s} r_k \right)^{-s} \prod_{0\leq k' < k \leq 2s} (r_{k'}-r_k) \, ,
\label{detvdm}
\ee
and it obviously vanishes if and only if two numbers $r_k$ and $r_{k'}$, say, are equal. 
However, $ r$ being nonzero, one has $r^k/r^{k'} = r^{2(k-k')} \neq 0$ if $k \neq k'$, and the inverse matrices ${\sf M}_m^{-1}$ do exist for this choice of $(2s+1)^2$ directions $\vec{n}_{qr}$. An {\em explicit} form of the inverse of matrices such as ${\sf M}_m$ has been worked out in {\cite{newton+68}.  

The reconstruction of pure states in systems with half-integer spin $s$ proceeds  similarly. However, a nonzero shift is necessary: $\Delta= 1/(2s+1)$, say. In this case, one obtains again matrices ${\sf M}_m$ ($0\leq m \leq s+1/2$) of type Vandermonde: for $0\leq k \leq 2s-m$ the entries simply read $({\sf M}_m)_{qk} = (r_k)^{s-q}$, while the remaining entries are given by
\be
({\sf M}_m)_{qk} =   \, \left( r_k \exp [-i 2 \pi \Delta]                                                           \right)^{s-q} \exp [i2\pi s \Delta]\, , 
                       \qquad 2s-m+1\leq k \leq 2s\, .
\label{phasedelem}
\ee
These extra phases $\exp [-i 2 \pi \Delta]$ distinguish lines which otherwise would be identical, a coincidence which does not occur for integer values of $s$. Therefore, all the matrices ${\sf M}_m$ ($0\leq m \leq s+1/2$) are invertible, too. For a spin $s=1/2$, the directions $\vec{n}_{00}$ and $\vec{n}_{01}$ are located in the $yz$ plane, while $\vec{n}_{10}$ and $\vec{n}_{11}$ are in the $xz$ plane, providing thus four independent numbers to determine the (unnormalized) density operator. With a zero shift $\Delta$, one would obtain information related to the $yz$ plane only which is {\em not} sufficient for reconstruction.      

In summary, it is possible to reconstruct the density matrix $\rho$ from $(2s+1)^2$ probabilities $p_s(\vec{n}_{qr})$, $1 \leq q,r \leq 2s+1$ along judiciously chosen directions $\vec{n}_{qr}$ using a Fourier transform and standard linear algebra. The Fourier transform reduces the original problem of inverting a $(2s+1)^2 \times (2s+1)^2$ matrix to that of inverting a number of $(2s+1) \times (2s+1)$ matrices of Vandermonde type. The method presented here is an {\em optimal} tool for reconstructing 
a spin state in the sense that a minimal number of data has to be acquired using nothing but a standard Stern-Gerlach apparatus.    
\section*{Acknowledgements}
St. W. acknowledges financial support by the {\em Schweizerische Nationalfonds}.  
\subsection*{Appendix}
In \cite{amiet+98/1},  four different approaches have been proposed to determine the coefficients $\rho^{*}_{lm}$, defined by expanding the statistical operator $\rho$ for a spin $s$ according to $\rho = 1/(2s+1) \sum_{lm}\rho_{lm}^{*}K_{lm}$, where the $K_{lm},$ $l=0,\ldots, 2s,$ $-l\leq m \leq l$, are a set of $(2s+1)^2$ orthogonal multipole operators \cite{fano+59}. The reconstruction of the statistical operator $\rho$ is based throughout on the inversion of the relation
\begin{equation}
\Pi_l(\theta,\varphi) 
       = \left( \frac{4\pi}{2l+1} \right)^{\frac{1}{2}} 
         \sum_{m=-l}^l Y_{lm}(\theta,\varphi) \rho^{*}_{lm} \, ,
\label{fundamentalrel}
\end{equation}
where the functions $Y_{lm}(\theta,\varphi)$ are standard spherical harmonics. The functions on the left-hand-side of this equation are linear combinations of measurable quantities,\footnote{The sums in Eqs.\ (11) and (13) in \cite{amiet+98/1} both go from $-s$ to $s$, not from $-l$ to $l$.}  
\begin{equation}
\Pi_l(\theta,\varphi) 
       = \sqrt{2s+1} \sum _{\mu=-s}^s 
        (-1)^{s-\mu}(s\mu,s-\mu|l0) p_{\mu}(\theta,\varphi) \, ,
\label{linearcomb}
\end{equation}
where $(s \mu,s \mu'|lm)$ denotes a Clebsch-Gordan coefficient, and $p_{\mu}(\theta,\varphi)$ is the probability to measure the value $\mu$ if the orientation of the Stern-Gerlach apparatus defines  ${\vec n}(\theta, \varphi)$ as axis of quantization. 

Unfortunately, the fourth method to invert Eq.\ (\ref{fundamentalrel}) is erroneous.\footnote{The authors would like to thank R.\ F.\ Werner for pointing this out.} It has been proposed to measure the probabilities at a fixed angle $\theta_j=\theta_M$, and angles $\varphi_k=k\,2\pi/(2s+1)$, $k=0,1,\ldots, 2s$, corresponding to $(2s+1)$ directions located on a cone about the $z$ axis. Contrary to what is stated just before Eq. (22) in \cite{amiet+98/1}, no orthogonality relation is available for the the angles $\varphi_k$; instead one has 
\begin{equation}
\frac{1}{2s+1} \sum_{k=0}^{2s} \exp [i(m-m')\varphi_k] 
        = \delta_{m\, m'} + \delta_{m\, m'+(2s+1)} + \delta_{m\, m'-(2s+1)}
\label{correctdeltas}
\end{equation}
since $-4s\leq m-m'\leq 4s$ (as follows from $-2s \leq m, m' \leq 2s$ and not $-s \leq m,m' \leq s$, as stated incorrectly), and the inversion of (\ref{fundamentalrel}) becomes impossible. Therefore, knowing the values $\Pi_l (\theta_M, \varphi_k)$ is {\em not} sufficient to determine the coefficients $\rho^{*}_{lm}$ unambiguously. One way out of this difficulty is to measure the probabilities along $(4s+1)$ (instead of $(2s+1)$) directions distributed homogeneously on a cone. Then, relation (\ref{correctdeltas}) is replaced by       
\begin{equation}
\frac{1}{4s+1} \sum_{k=0}^{4s} \exp [i(m-m')k 2 \pi/(4s+1) ] 
        = \delta_{m \, m'} \, ,
\label{toomany}
\end{equation}
and the intended inversion becomes possible. However, the number of measured parameters 
has been increased considerably. Effectively, one obtains a method of state reconstruction which is equivalent to the approach developed in {\cite{newton+68}.
\end{document}